\documentclass[a4paper,11pt]{article}
\pdfoutput=1 % if your are submitting a pdflatex (i.e. if you have
\usepackage{jheppub} % for details on the use of the package, please
\usepackage[T1]{fontenc} % if needed
\usepackage{psfrag}
\usepackage[utf8]{inputenc} 
\usepackage{hyperref}
\usepackage{caption}% oder capt-of, oder ohne die Zeile mit KOMA-Script-Klasse

\title{\boldmath Colorful vortex intersections in SU($2$) lattice gauge theory and their influencs on chiral properties.}

\author[a]{Seyed Mohsen Hosseini Nejad}
\author[b]{and Manfried Faber}

\affiliation[a]{Department of Physics, University of Tehran, P.O. Box 14395-547, Tehran 1439955961, Iran}
\affiliation[b]{Institute of Atomic and Subatomic Physics, Vienna University of Technology, Wiedner Hauptstr. 8-10, 1040 Vienna, Austria}

\emailAdd{smhosseininejad@ut.ac.ir}
\emailAdd{faber@kph.tuwien.ac.at}

\abstract{We introduce topological non-trivial colorful regions around intersection points of two perpendicular vortex pairs and investigate their influence on topological charge density and eigenmodes of the Dirac operator. With increasing distance between the vortices the eigenvalues of the lowest modes decrease. We show that the maxima and minima of the chiral densities of the low modes follow mainly the distributions of the topological charge densities. The topological non-trivial color structures lead in some low modes to distinct peaks in the chiral densities. The other low modes reflect the topological charge densities of the intersection points.}


\def\gtwid{{\,\raise.3ex\hbox{$>$\kern-.75em\lower1ex\hbox{$\sim$}}\,}}

\begin{document} 
\maketitle
\flushbottom
%------------------------------------------------------------------------------
\section{Introduction}

Non-perturbative QCD is dominated by the phenomena of color confinement and spontaneous chiral symmetry breaking ($\chi$SB). The non-perturbative nature of these phenomena is caused by the non-triviality of the QCD vacuum. There is still intensive discussion about the nature of the quantum fluctuations responsible for this non-triviality. The idea that the QCD vacuum is dominated by center vortices~\cite{tHooft:1977nqb,vinciarelli:1978kp,yoneya:1978dt,cornwall:1979hz,mack:1978rq,nielsen:1979xu}, by quantized magnetic flux tubes, is strongly supported by lattice simulations~\cite{deldebbio:1996mh,langfeld:1997jx,deldebbio:1997ke,langfeld:1998cz,kovacs:1998xm,bertle:2002mm} and infrared models~\cite{engelhardt:1999wr,engelhardt:1999fd,engelhardt:2003wm,hollwieser:2014lxa,altarawneh:2015bya,hollwieser:2015qea,altarawneh:2016ped}. Via the Stokes theorem vortices piercing Wilson loops modify the trace of these loops by center elements. A non-vanishing density of center vortices causes an exponential decay of Wilson loops proportional to the minimal area of the loops and a confinement potential getting linear in the infrared. Vortex removal destroys confinement and restores chiral symmetry~\cite{deforcrand:1999ms,alexandrou:1999vx}. This observation encourages the conclusion that vortices are responsible for $\chi$SB, but it does not yet explain how vortices may lead to this phenomenon.

According to the Banks-Casher relation~\cite{banks:1979yr}, the density of near-zero modes of the Dirac operator is proportional to the value of the chiral condensate, the order parameter of $\chi$SB. The Atiyah-Singer index theorem~\cite{atiyah:1971rm,brown:1977bj,adams:2000rn} relates the difference in the numbers of zero-modes of positive and negative chirality to the topological charge of gauge field configurations. A well established theory of $\chi$SB relies on instantons~\cite{belavin:1975fg,actor:1979in,tHooft:1976snw,bernard:1979qt}, which are localized in space-time and carry a topological charge of modulus one. A zero mode of the Dirac operator arises, which is concentrated at the instanton core. The instanton liquid model~\cite{ilgenfritz:1980vj,diakonov:1984vw,diakonov:1986tv} provides a mechanism how overlapping would-be zero modes split into low-lying nonzero modes which create a chiral condensate. But instantons are minima of the action and therefore only of indirect importance in the ensemble of gauge field configurations as far as they increase the number of local minima of the action and all lumps of topological charge can finally be deformed via cooling or smoothing procedures to instantons. Since smooth deformations of field configurations do not change the homotopy class, these observations indicate that all types of contributions to the topological charge may influence the density of near-zero modes via their interaction.

Lattice simulations have shown that center vortices contribute to the topological charge via writhing, vortex intersections~\cite{engelhardt:2000wc,bertle:2001xd,bruckmann:2003yd,engelhardt:2010ft,hollwieser:2011uj,hollwieser:2014mxa,nejad:2015aia,hollwieser:2015koa} and their color structure~\cite{jordan:2007ff,hollwieser:2010zz,hollwieser:2012kb,schweigler:2012ae}. Vortices lead also to spontaneous $\chi$SB~\cite{engelhardt:1999xw,reinhardt:2000ck,engelhardt:2002qs,leinweber:2006zq,bornyakov:2007fz,hollwieser:2008tq,bowman:2010zr,hollwieser:2013xja,brambilla:2014jmp,hoellwieser:2014isa,trewartha:2014ona,trewartha:2015nna}. As Engelhardt and Reinhardt~\cite{engelhardt:1999xw,engelhardt:2000wc} have indicated in SU(2) gauge theory in addition to the location of the vortex surfaces, one needs their orientation to determine contributions of vortices to the topological charge. This relates to the common vortex identification method on the lattice, resulting in P-vortices~\cite{deldebbio:1996mh}, thin connected surfaces consisting of plaquettes projected to center elements. Regions of different orientations are separated by lines, attributed to the worldlines of magnetic monopoles in maximal abelian gauge~\cite{ambjorn:1999ym}. In a Monte-Carlo ensemble of field configuration vortices are neither thin nor do they have a given orientation, they have a certain average thickness and colors of the full gauge group. At present the determination of the color structure of vortices is impossible. Therefore, the only possibility to study the influence of the color structure of thick vortices on the topological charge density and low lying eigenmodes of the Dirac operator is by studies of artificial gauge configurations. In recent articles the importance of colorful vortices has been underlined~\cite{schweigler:2012ae,nejad:2015aia}. The contributions of intersections with such colorful vortices has not been investigated yet and is at the focus of this article.

We restrict our investigation to SU(2) lattice gauge theory. On periodic lattices plane vortices have to appear in parallel or anti-parallel pairs. We arrange two such pairs appropriately to get intersections in four points, as discussed in Ref.~\cite{hollwieser:2013xja} where the intersections of uni-color vortices, of vortices in a U(1)-subgroup are investigated. In this paper we modify such a configuration and follow the suggestion of Ref.~\cite{nejad:2015aia} and make one of the plane vortices colorful, this means the links of this colorful vortex are distributed over the full SU(2) gauge group. After a gauge transformation it gets obvious that a colorful vortex is a vacuum to vacuum transition along a direction perpendicular to the vortex.

In Sect.~\ref{Sect1} we describe the gluon field configurations investigated in this article: two perpendicular plain anti-parallel vortex pairs intersecting in four points, where one of the vortices is colorful around one of the intersections. We investigate the influence of the position of the colorful region on the topological charge $Q$ in Sect.~\ref{Sect2}. There we check also the details of the topological charge density and give an interpretation of its behavior. In Sect.~\ref{Sect3} we analyze the low lying modes of the overlap Dirac operator \cite{narayanan:1993ss,narayanan:1994gw,neuberger:1997fp,edwards:1998yw} in the background of the considered vortex configurations. We study the influence of the distances of the vortex pairs on the lowest eigenvalues of the overlap Dirac operator. We determine the chiral densities of zero modes and near-zero modes and compare them with excited modes. We find pronounced peaks and regions with oscillating chiral density and try to find correlations of chiral and topological charge densities.

%------------------------------------------------------------------------------
\section{Uni-color  and Colorful SU($2$) plane vortices }\label{Sect1}

\begin{figure}[h!]
\centering
a)\includegraphics[width=0.36\columnwidth]{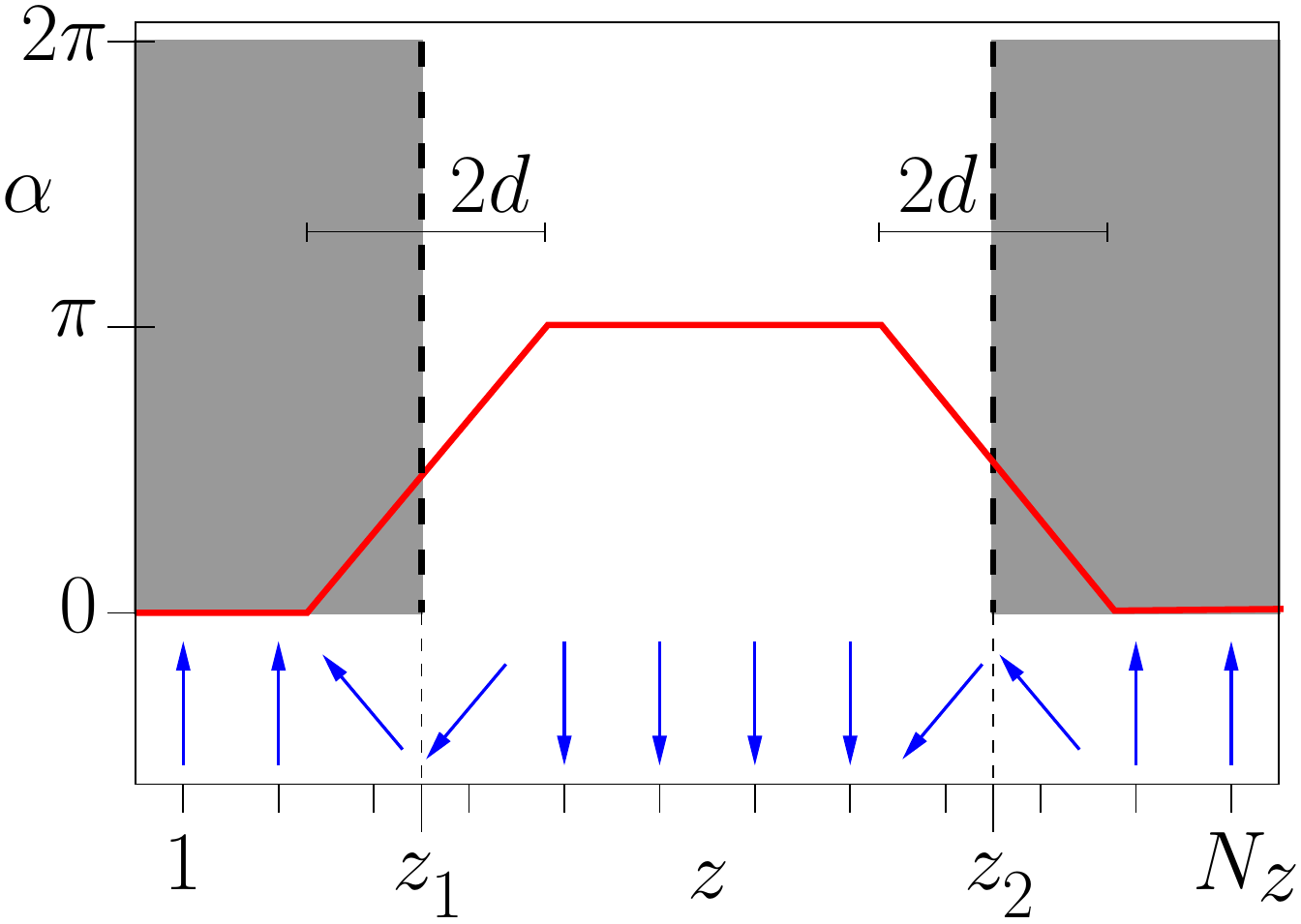}
b)\includegraphics[width=0.26\columnwidth]{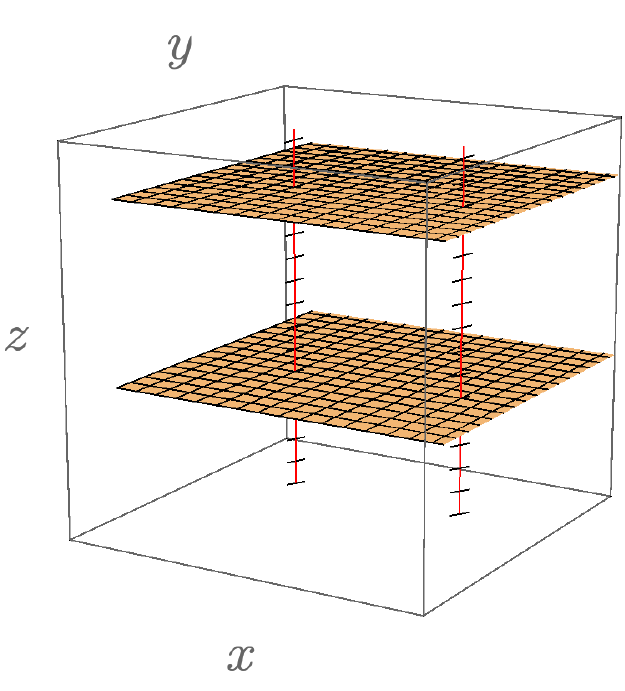}
c)\includegraphics[width=0.26\columnwidth]{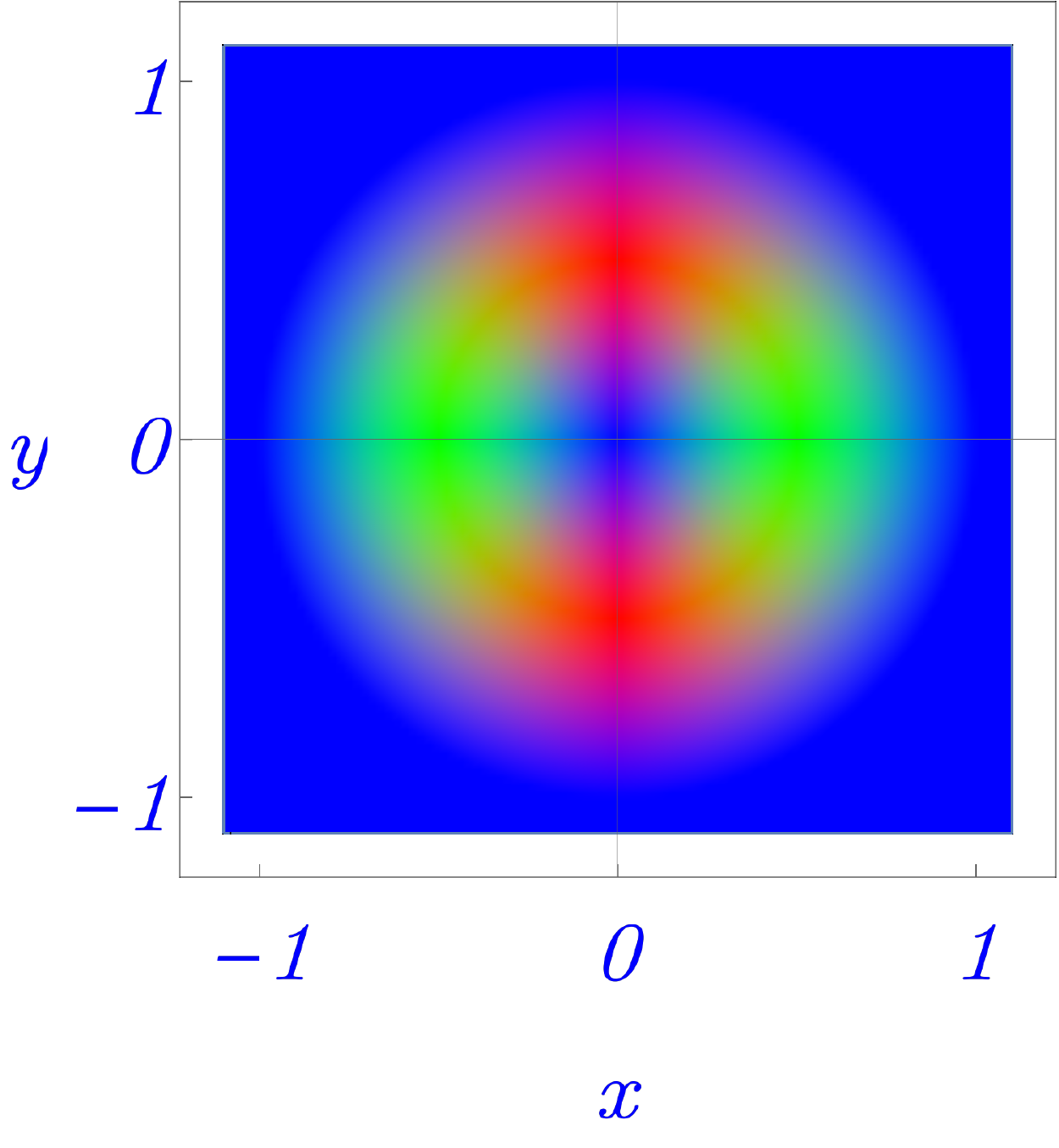}
\caption{a) The profile function $\alpha(z)$ of an anti-parallel $xy$ vortex pair with t-links varying in z-direction. The arrows indicate t-links rotating in $z$ direction. The centers of the thick vortices (dashed lines) are located at $z_1$ and $z_2$. In the shaded areas the links have positive, otherwise negative trace. b) Three-dimensional detail of two pairs of intersecting vortices in 4D. The horizontal planes represent xy-vortices and the vertical lines a t-slice of zt-vortices. The two pairs intersect in four points of a yt-plane. c) Diagram of a colorful xy-vortex. The color direction $\vec n$ of t-links is displayed in the xy-plane for $R=1$ and $x_0=y_0=0$ by maps to RGB-colors, $\pm\hat i\mapsto$ green, $\pm\hat j\mapsto$ red and $\pm\hat k\mapsto$ blue.}
\label{fig:pgc}
\end{figure}

The configurations which we want to investigate in SU($2$) lattice gauge theory are thick plane vortices \cite{jordan:2007ff,hollwieser:2011uj} extending along two coordinate axes, thickness in a third coordinate direction and formulated with non-trivial links in the forth direction. One of these vortices will get a special color structure and will be smoothed in the forth direction. On periodic lattices due to their quantization with the non-trivial center element, plane vortices have to occur in pairs. We use two different arrangements of vortex sheets, xy-vortices formulated with t-links in a given t-slice $t_\perp$ changing in $z$-direction and zt-vortices with non-trivial y-links at the y-slice $y_\perp$ varying in $x$-direction. For uni-color vortices the nontrivial links are elements in a U($1$) subgroup of SU($2$), usually the $\sigma_3$-subgroup with links of the form
\begin{equation}\label{def_rn}
U_\mu(x)=\exp\{\mathrm i\alpha(x)\sigma_3\}.
\end{equation}
To such U(1)-vortices we can assign an orientation given by the gradient of the angle $\alpha$. We treat pairs of anti-parallel plane vortices where in a region of $2d$ given by the vortex thickness the angle $\alpha$ increases in the first vortex linearly from 0 to $\pi$ and decreases in the next vortex linearly from $\pi$ to 0. We define the profile function $\alpha(z)$ for a pair of xy-vortices as
\begin{equation}\label{eq:phi-pl0}
\alpha(z)=\begin{cases}0&0<z\leq z_1-d,\\ 
                \frac{\pi}{2d}[z-(z_1-d)]&z_1-d< z\leq z_1+d,\\ 
                \pi&z_1+d<z\leq z_2-d,\\ 
                \pi\left[1-\frac{z-(z_2-d)}{2d}\right]&z_2-d<z\leq z_2+d,\\ 
                0&z_2+d<z\leq N_z.\end{cases}
\end{equation}
A diagram for such a profile function is shown in Fig.~\ref{fig:pgc}a, see also ref~\cite{hollwieser:2011uj}. An analog profile we use for $\alpha(x)$ for a pair of anti-parallel zt-vortices centered around $x_1$ and $x_2$. The two vortex pairs intersect in the $y_\perp,t_\perp$-plane at four points with the coordinates $x_1,x_2$ and $z_1,z_2$. The intersection of such vortex pairs is displayed in Fig.~\ref{fig:pgc}b.

There is a two-dimensional manifold of U(1)-subgroups of SU(2). These subgroups can be characterized by a unit vector $\vec n$ in the SU(2) group element $\exp\{\mathrm i\alpha\vec n\vec\sigma\}$  defining a unit sphere S$^2$ in $\mathbb R^3$. Mapping this S$^2$ to a two-dimensional vortex plane leads to a colorful vortex as introduced in Ref.~\cite{nejad:2015aia}. We map this S$^2$ to the time-like links of a circular region with radius $R$ around $x_0,y_0$ of the xy-vortex at $z_1,t_\perp$ by
\begin{equation}\begin{aligned}\label{DefColVort}
&\vec n\vec\sigma=\sigma_1\,\sin\theta(\rho)\cos\phi(x,y)
+\sigma_2\,\sin\theta(\rho)\sin\phi(x,y)+\sigma_3\,\cos\theta(\rho),\\
&\rho=\sqrt{(x-x_0)^2+(y-y_0)^2},\\
&\theta(\rho)=\pi(1-\frac{\rho}{R})H(R-\rho)\;\in\,[0,\pi],\quad
\phi=\arctan_2\frac{y-y_0}{x-x_0}\;\in\,[0,2\pi),
\end{aligned}\end{equation}
where $H$ is the Heaviside step function. The color structure of such a vortex is displayed in Fig.~\ref{fig:pgc}c. As discussed in Refs.~\cite{schweigler:2012ae,nejad:2015aia} colorful vortices defined by links in one time-slice of the lattice do not contribute to $\vec E_a\vec B_a$ and have vanishing gluonic topological charge. The sum over the index $a$ runs over the 3 directions $\sigma_a$ of the SU(2) color algebra. The vanishing of this contribution is a lattice artifact due to the singularity of the vortex in time direction.  By a gauge transformation rotating the non-trivial time-like links to unit matrices it gets obvious that a colorful plane vortex defines a transition between vacua of different winding number. After this gauge transformation it is possible to smooth the vortex in time direction without creating a singularity of the gauge field~\cite{schweigler:2012ae,nejad:2015aia}. Increasing the smoothing region $\Delta t$ of the colorful vortex described in Eq.~(\ref{DefColVort}) the gluonic topological charge approaches $Q=-1$, as will be also described in Sect.~\ref{Sect2}.

After defining uni-color and colorful vortices we are ready to compare the topological charge contributions of colorful intersections with those of uni-color intersections.

%------------------------------------------------------------------------------
\section{Topological Charges and their Densities}\label{Sect2}

According to the definition
\begin{gather}\label{eq:qlatq}
  Q=-\frac{1}{32\pi^2}\int d^4x\,\epsilon_{\mu\nu\rho\sigma}\,
\mbox{tr}[{\mathcal F}_{\mu\nu}{\mathcal F}_{\rho\sigma}]
=\frac{1}{4\pi^2}\int d^4x\,\vec E_a\cdot\vec B_a
\end{gather}
only regions with common presence of electric and magnetic fields of same spatial directions and colors $a$ contribute to the topological charge $Q$. The configurations introduced in Sect.~\ref{Sect1}, two perpendicular plain anti-parallel vortex pairs, contribute at vortex intersections and in colorful regions. We intersect an anti-parallel $xy$- with an anti-parallel zt-vortex pair, as shown in Fig.~\ref{fig:pgc}b. For uni-color vortices each intersection point gives rise to a lump of topological charge $Q=\pm1/2$ \cite{reinhardt:2002cm}. For anti-parallel vortex pairs two of the intersection points carry topological charge $Q=+1/2$ while the other two intersection points have $Q=-1/2$ \cite{hollwieser:2011uj}. They sum up to total topological charge $Q=0$.

It is interesting to investigate the modification of these contributions when one of the uni-color vortices is substituted by a colorful vortex, with the colorful region centered at one of the intersection points. The continuum action $S$ for a colorful region with radius $R$ and smoothing region $\Delta t$ is calculated as \cite{nejad:2015aia}

\begin{equation}
\frac{S(\Delta t)}{S_\mathrm{Inst}}=\frac{0.51\,\Delta t}{R}
+\frac{1.37\,R}{\Delta t}
\end{equation}
where the instanton action $S_\mathrm{Inst}=8\pi^2/g^2$. Its minimum value is reached around $R=\Delta t$ with 1.68~$S_\textrm{Inst}$. 

\begin{figure}[h!]
\centering
a)\includegraphics[width=0.46\columnwidth]{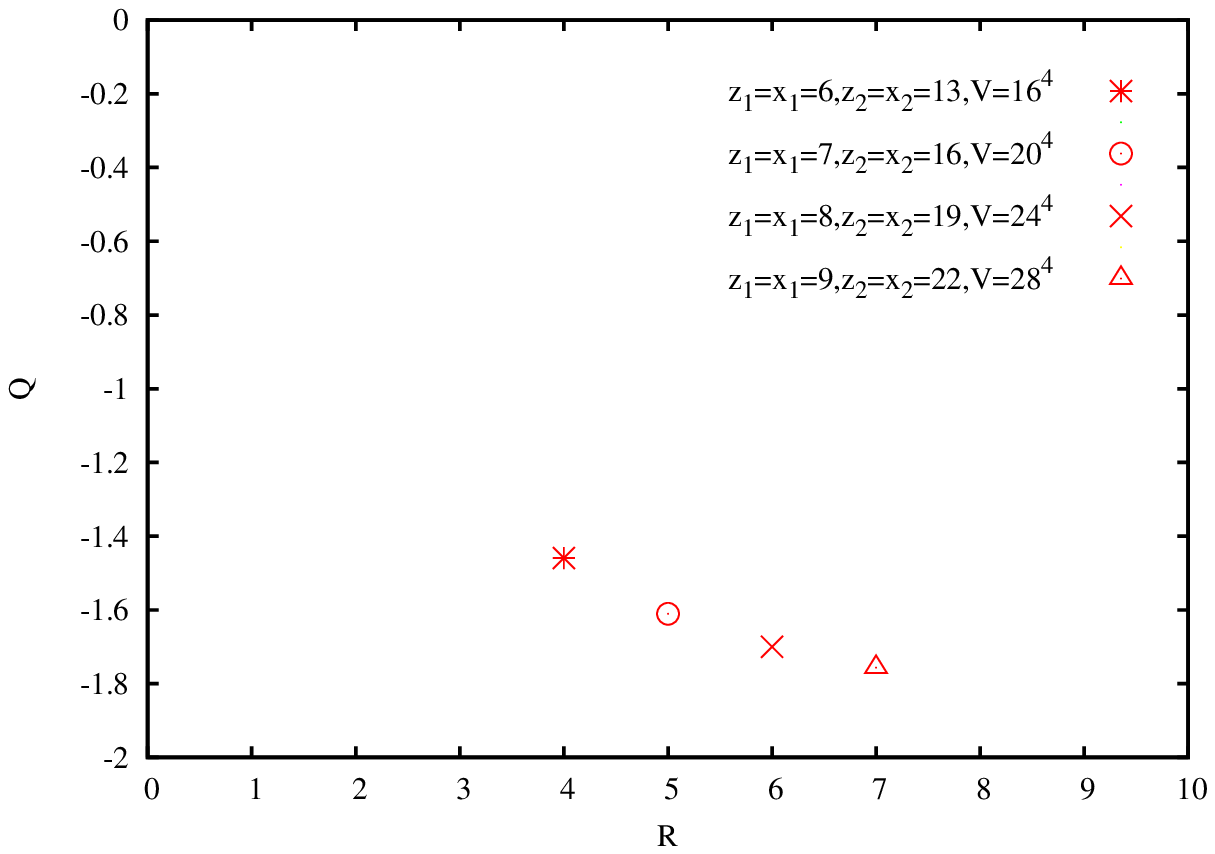}
b)\includegraphics[width=0.46\columnwidth]{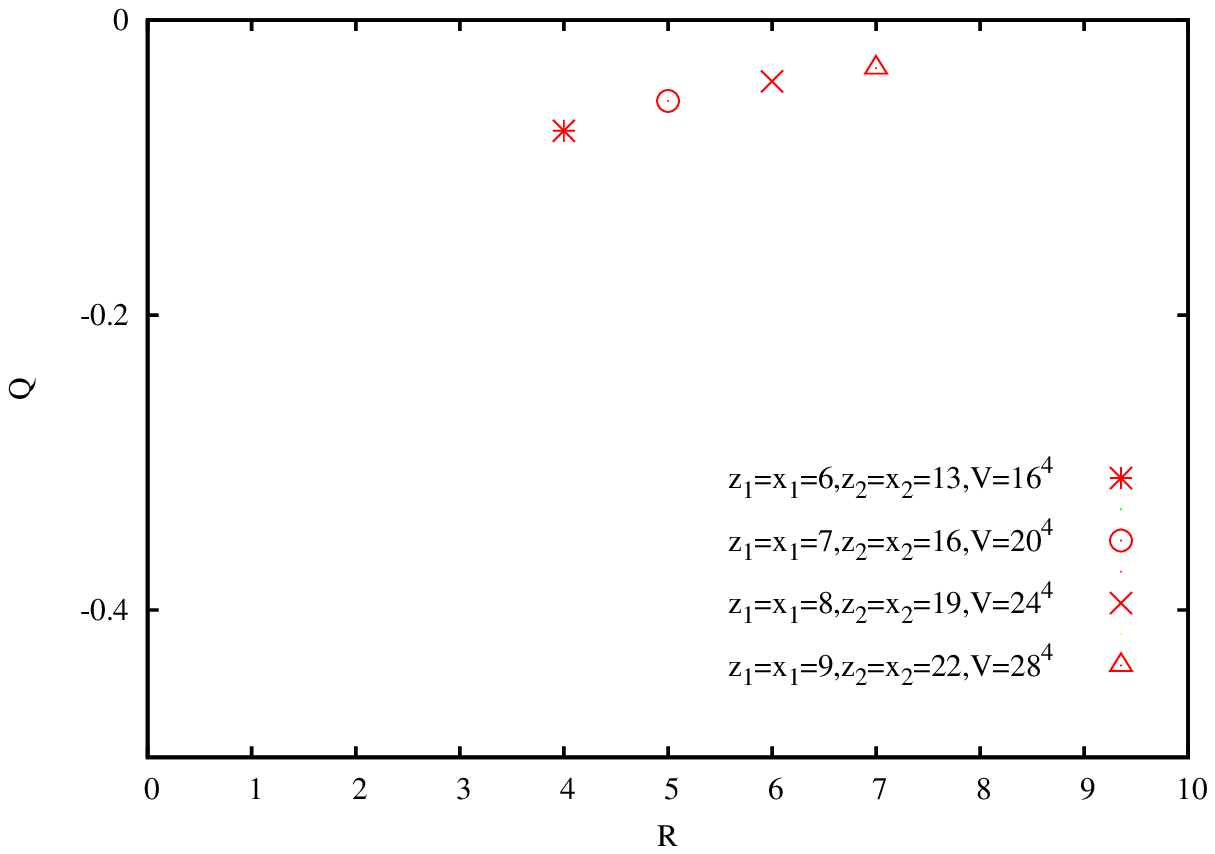}
\caption{a) The total topological charge of the vortex configurations corresponding to Fig.~\ref{fig:schematic}a and \ref{fig:schematic}b. In the left (right) diagram the colorful region in the xy-plane is centered at $x_0=x_1\;(x_0=x_2)$ and $y_0=y_\perp$. Increasing the radius $R$ of the colorful region and increasing the lattice size the total topological charge converges to $Q=-2$ for the configuration of Fig.~\ref{fig:schematic}a and to $Q=0$ for Fig.~\ref{fig:schematic}b.}
\label{fig:q1}
\end{figure}
\begin{figure}[h!]
\centering
a)\includegraphics[width=0.46\columnwidth]{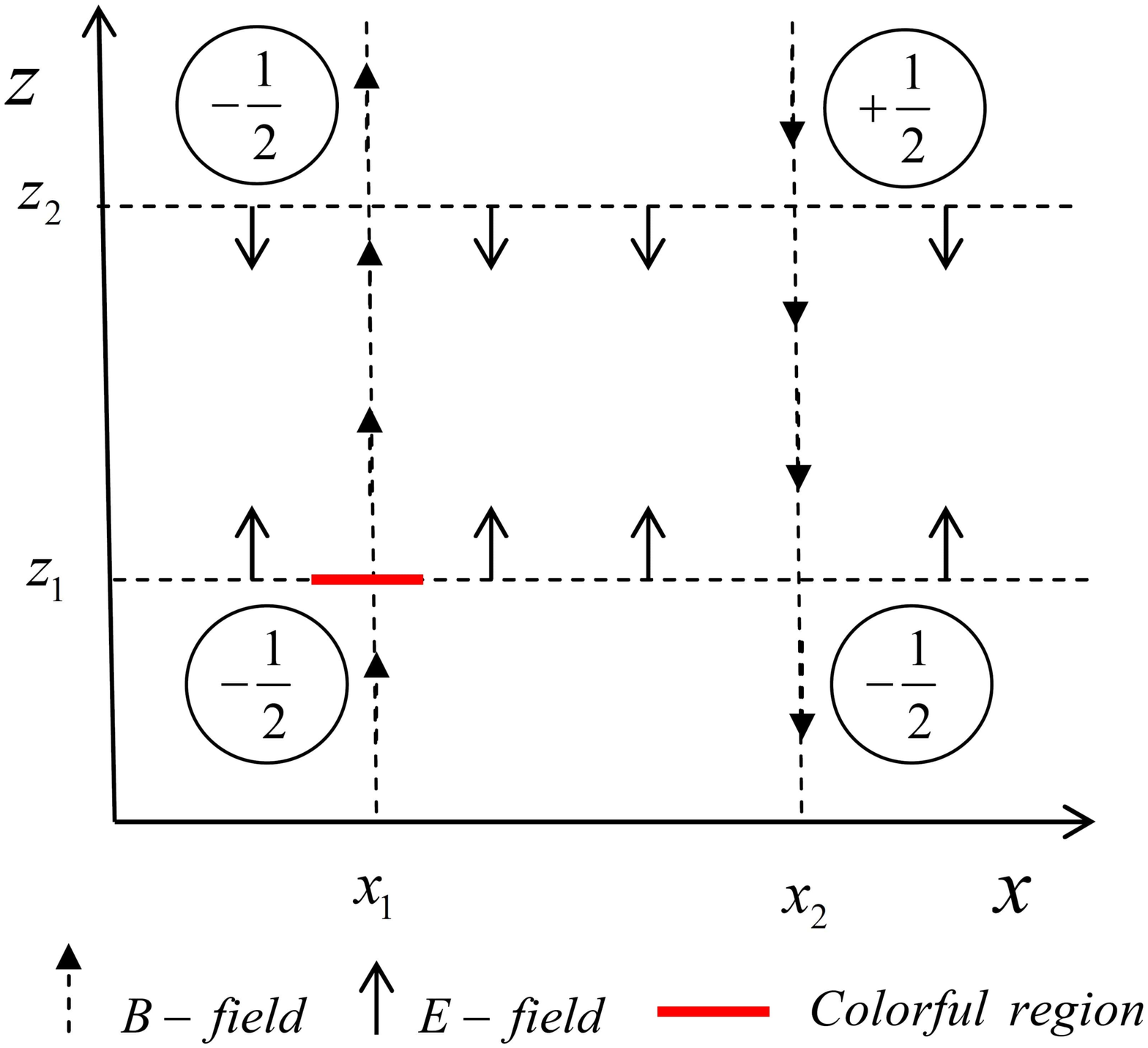}
b)\includegraphics[width=0.46\columnwidth]{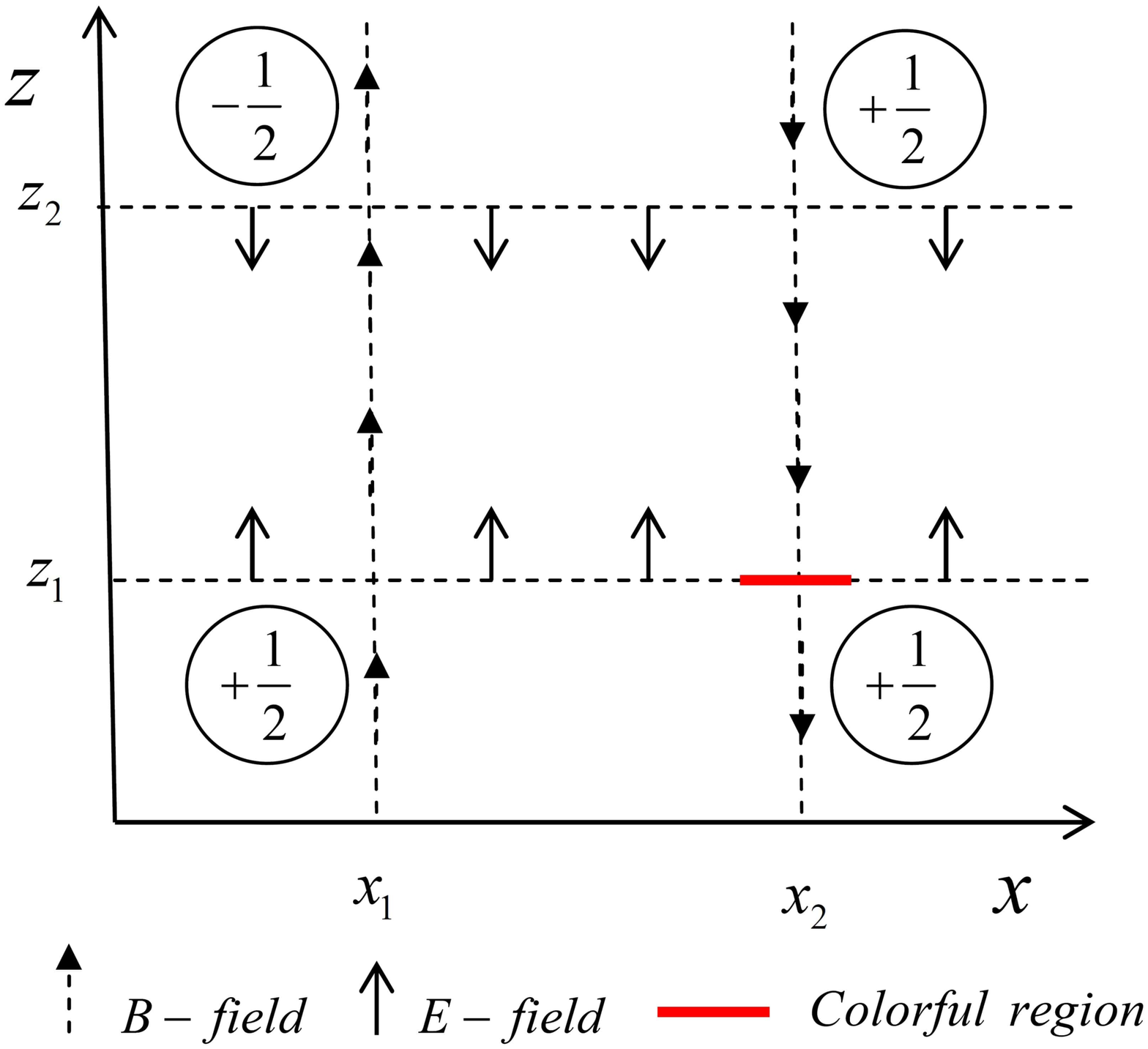}
\caption{The geometry, field strength and the contribution to the topological charge of the intersection regions in the intersection plane of  two anti-parallel vortex pairs. In diagram (a) the red line at the intersection point $(x_1,z_1)$ indicates that the uni-color vortex is in this region substituted by a colorful region. In the right diagram (b) the region around $(x_2,z_1)$ is substituted.}
\label{fig:schematic}
\end{figure}

The total topological charge of the configuration with the colorful region around $(x_1,z_1)$ is shown in Fig.~\ref{fig:q1}a and with the colorful region around  $(x_2,z_1)$ in Fig.~\ref{fig:q1}b for $\Delta t=R$ for various values of $R$ and increasing lattice sizes. The values of the topological charge approach the values $Q=$-2 and 0. To explain these asymptotic values we display in Fig.~\ref{fig:schematic} schematic diagrams for the intersection planes of these configurations.
These diagrams show that by the insertion of the colorful regions the topological charge contributions at the intersection points change sign leading to sums of their four contributions of $\mp1$. Such a sign change corresponds to the change of orientation of the corresponding uni-color vortex within a circle of radius $R$, by the insertion of a circular monopole line around the intersection point. But in the case of a colorful vortex it is a monopole line which changes its color along the circle in a non-trivial way, such that this line contributes itself with a value of -1 to the total topological charge $Q=\mp1-1$ of the two investigated configurations.

\begin{figure}[h!]
\centering
a)\includegraphics[scale=0.5]{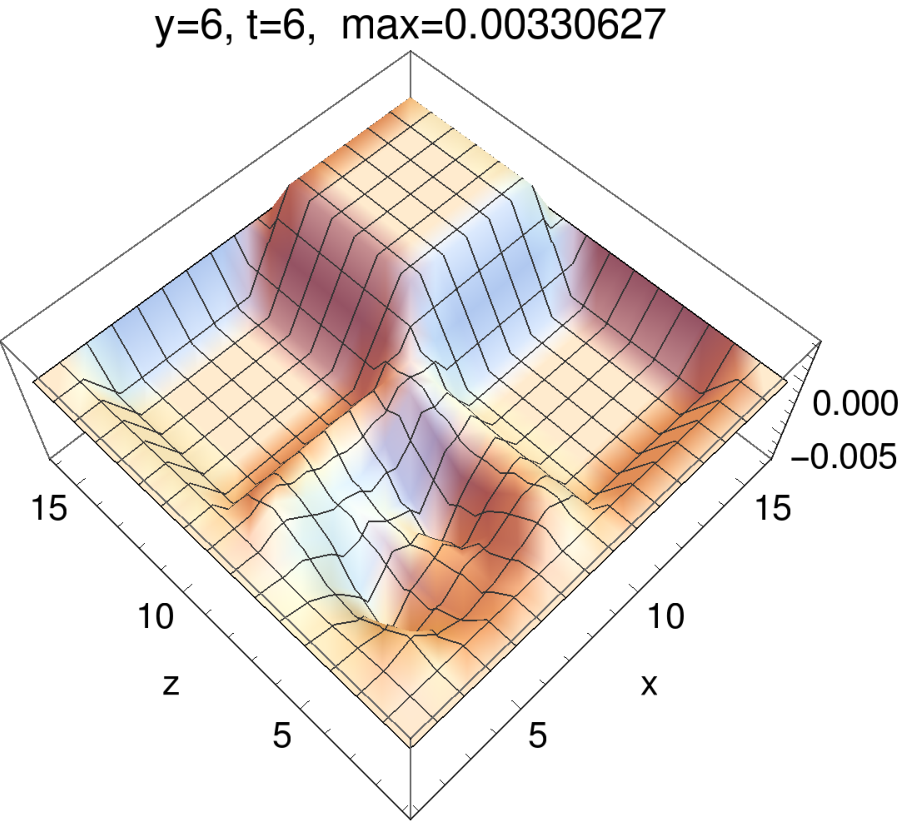}
b)\includegraphics[scale=0.5]{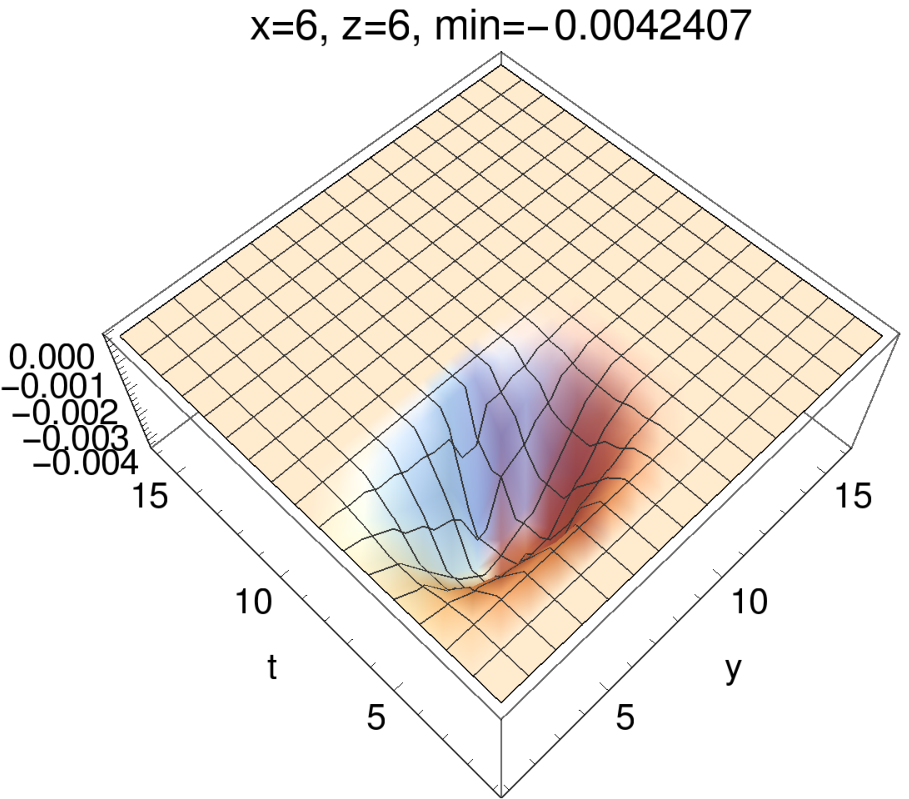}
c)\includegraphics[scale=0.5]{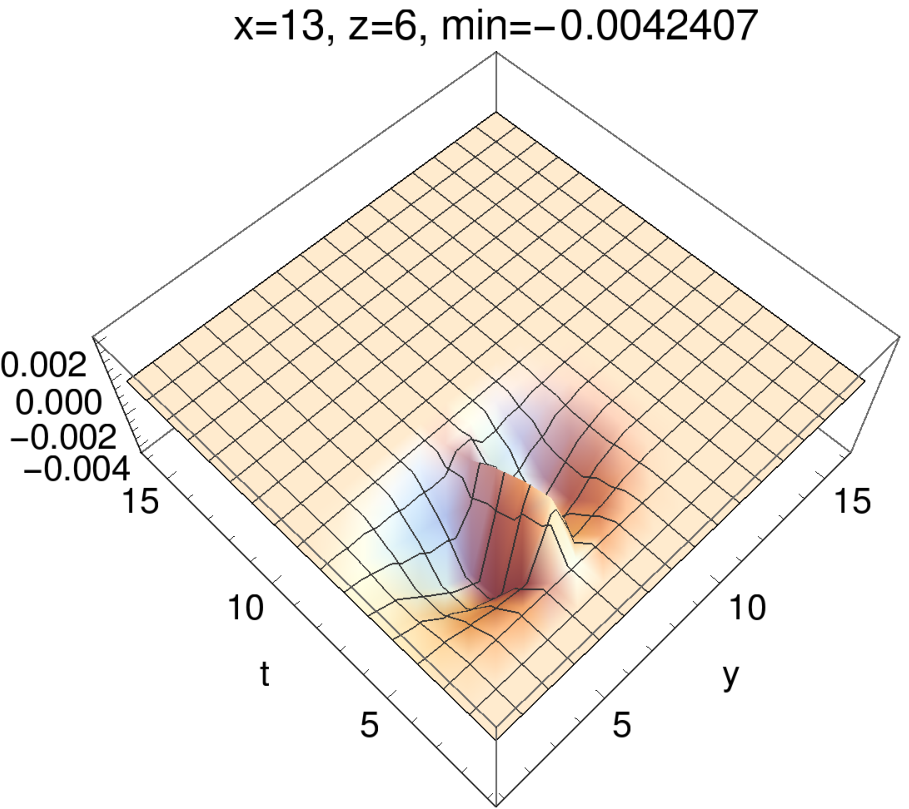}
\caption{The topological charge density in two characteristic planes for two intersecting anti-parallel  xy- and zt-vortex pairs with ($z_1=6,z_2=13$) and ($x_1=6,x_2=13$) at $t_\perp=y_\perp=6$ with thickness $d=3$ and $\Delta t=R=4$ on a $16^4$-lattice is displayed in the diagrams a) and b) for the situation of Fig.~\ref{fig:schematic}a, where the center of the colorful region of the  xy-vortex surrounds the intersection at lower $x$ and lower $z$, $x_0=x_1$ and $z_0=z_1$. Diagram a) shows the topological charge density in the intersection plane $y_\perp=t_\perp=6$ with the four intersection regions. The area around the intersection point $x_1=z_1=6$ is influenced by the surrounding color structure shown in the perpendicular yt-plane in diagram b) as a broad, shallow depression. The intersection region at $x_1=z_1=6$ forms a narrow, deep structure since the zt-vortex is defined by y-links in one y-slice only. For the situation of Fig.~\ref{fig:schematic}b the two contributions at the intersection point $x_0=x_2$ and $z_0=z_1$, a broad structure from the colorful region and a narrow one from the intersection, can be better distinguished in diagram c), due the difference in their signs. Diagram c) displays the topological charge density in the yt-plane at the intersection point $(x_2,z_1)$ for the situation of Fig.~\ref{fig:schematic}b, where now this other intersection is surrounded by the colorful region.}
\label{fig:topden}
\end{figure}

To check the details of the contribution to the topological charge we show in Figs.~\ref{fig:topden}a and \ref{fig:topden}b characteristic charge densities of an anti-parallel xy-vortex pair at ($z_1=6,z_2=13$) with an anti-parallel $zt$-pair at ($x_1=6,x_2=13$) at $t_\perp=y_\perp=6$ with thickness $d=3$ and $\Delta t=R=4$ on a $16^4$-lattice. In both diagrams, the center of the colorful region with radius $R = 4$ in the $xy$ plane is located at $x_0=x_1=6,\;y_0=y_\perp=6$. In Fig.~\ref{fig:topden}a the topological charge density of the vortex configuration is plotted in the intersection plane, the xz-plane with $y=y_\perp,t=t_\perp$, where we can identify the positive and negative contributions indicated in Fig.~\ref{fig:schematic}a. In Fig.~\ref{fig:topden}b we show the perpendicular yt-plane at the colorful intersection with the coordinates $x_1=z_1=6$. The broad shallower structure of the colorful vortex extends in y-direction with radius $R$ and in t-direction with $\pm\Delta t/2$ around $t_\perp$. There is a further contribution to the topological charge density from the intersection region of the xy-vortex with the zt-vortex. It is narrow in y since the zt-vortex is constructed from non-trivial y-links in one y-slice at $y_\perp$ only.

When the colorful region is shifted from $x_1$ to $x_2$ as schematically depicted in Fig.~\ref{fig:schematic}b the total topological charge gets $Q=0$. Since the orientation of the xy-vortex is now flipped at the intersection point $(x_2,z_1)$, the contribution of this intersection changes sign, $-\frac{1}{2}\to+\frac{1}{2}$, as indicated in Fig.~\ref{fig:schematic}b. Since the colorful region contributes again with $-1$ this results in the total topological charge $Q=0$. It is not difficult to imagine the analog of Fig.~\ref{fig:topden}a after this shift of the colorful region. Therefore, it is not shown by a diagram. But it may be instructive to see the analog of Fig.~\ref{fig:topden}b. It is displayed in Fig.~\ref{fig:topden}c. The shallower negative contribution reflects the contribution of the colorful vortex and the ridge in t-direction originates in the contribution of the vortex intersection which in this case is positive.

Eqs.~(\ref{eq:phi-pl0}) and (\ref{DefColVort}) can also be used to insert a colorful region with topological charge contribution of $+1$ instead of $-1$ which was discussed above. From Eq.~(\ref{eq:phi-pl0}) we can read that the gradient of $\alpha$ has opposite $z$-direction at $z_1$ and $z_2$. Thus, a shift of the center coordinate $z_0$ from $z_1$ to $z_2$ flips the sign of this gradient, but it leaves the $xy$-structure~(\ref{DefColVort}) untouched and leads therefore to a sign change of the topological charge of the colorful region. Combined with the contributions of the intersections, one of them modified by a surrounding colorful region, we get total topological charges $Q=0$ for $(x_0=x_1,z_0=z_2)$ and $Q=2$ for $(x_0=x_2,z_0=z_2)$. By symmetry considerations we can easily imply the consequences on the topological charge densities and also on eigenvalues of the Dirac operator and chiral densities which are discussed in the next section.

%------------------------------------------------------------------------------
\section{Dirac Eigenmodes and Chiral Densities}\label{Sect3}

In the previous section, we defined two colorful configurations which are combinations of two anti-parallel plane vortex pairs. Now, we investigate the effect of these configurations on fermions $\psi$ by determining the low-lying eigenvectors and eigenvalues $|\lambda| \in [0,2/a]$ of the overlap Dirac operator \cite{narayanan:1993ss,narayanan:1994gw,neuberger:1997fp,edwards:1998yw}
\begin{eqnarray}\label{eq:ov_dirac}
D_{ov}=\frac{1}{a}\left[1+ \gamma_5 \frac{H}{|H|}\right]
\textrm{ with }H=\gamma_5 A,\;A=a D_\mathrm{W}-m,
\end{eqnarray}
where $m$ describes one species of single massless Dirac fermions and has to be in the range $(0,2)$ and the massless Wilson Dirac operator $D_W$ \cite{wilson:1974sk,gattringer:2010zz} on a lattice with lattice constant $a$ reads
\begin{eqnarray}\label{WilsAct}
D_\mathrm{W}(x,y)=\frac{4}{a}\delta_{x,y}-
\frac{1}{2a}\sum_{\mu=\mp1}^{\pm4}(1-\gamma_\mu)\;U_\mu(x)\;\delta_{x+\hat\mu,y}
\textrm{ with }\gamma_{-\mu}=-\gamma_\mu,\;U_{-\mu}(x)=U_\mu^\dagger(x-\hat\mu).
\end{eqnarray}
The coordinates and the results for eigenvalues and densities we give in units of the lattice constant $a$, i.e we put further on $a=1$. The vectors $\hat\mu$ connect nearest neighbors $x$ and $y$ in $x_\mu$-direction. $U_\mu(x)\in SU(2)$ are the parallel transporters from $x$ to $x+\hat\mu$. The mass parameter $m$ is chosen with $m=+1.5$. The eigenvalues of the overlap Dirac operator as a Ginsparg-Wilson operator are restricted to a circle in the complex plane. The absolute value $|\lambda|$ of the two complex conjugate eigenvalues of $D_{ov}$ is simply written as $\lambda$. 
 
According to the Atiyah–Singer index theorem~\cite{atiyah:1971rm} a configuration with non-vanishing topological charge has to be related to zero modes of the Dirac operator. If lumps with topological charge are parts of a larger configuration they could localize the fermionic modes, interact and contribute to a finite density of near-zero modes. According to the Banks-Casher relation \cite{banks:1979yr} a finite density of near-zero modes leads to non-zero chiral condensate and spontaneous $\chi$SB. 

\begin{figure}[h!]
\centering
a)\includegraphics[width=0.46\columnwidth]{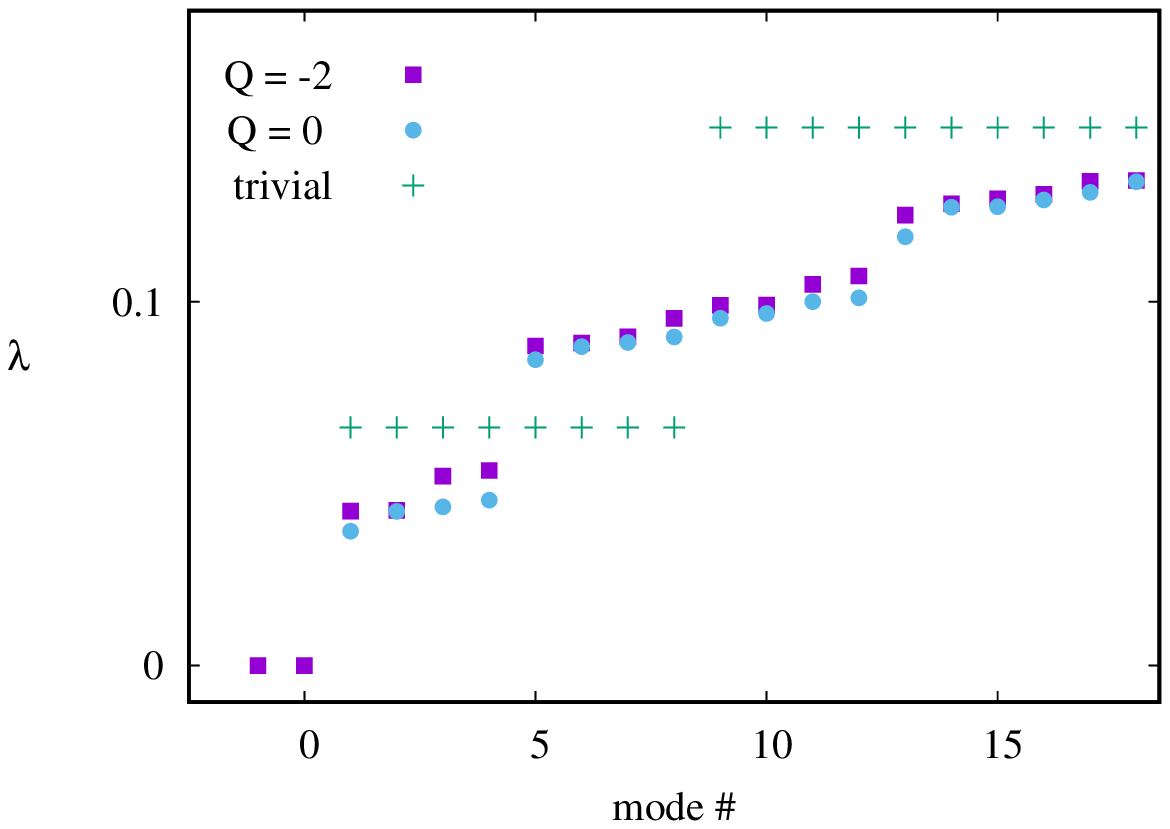}
b)\includegraphics[width=0.46\columnwidth]{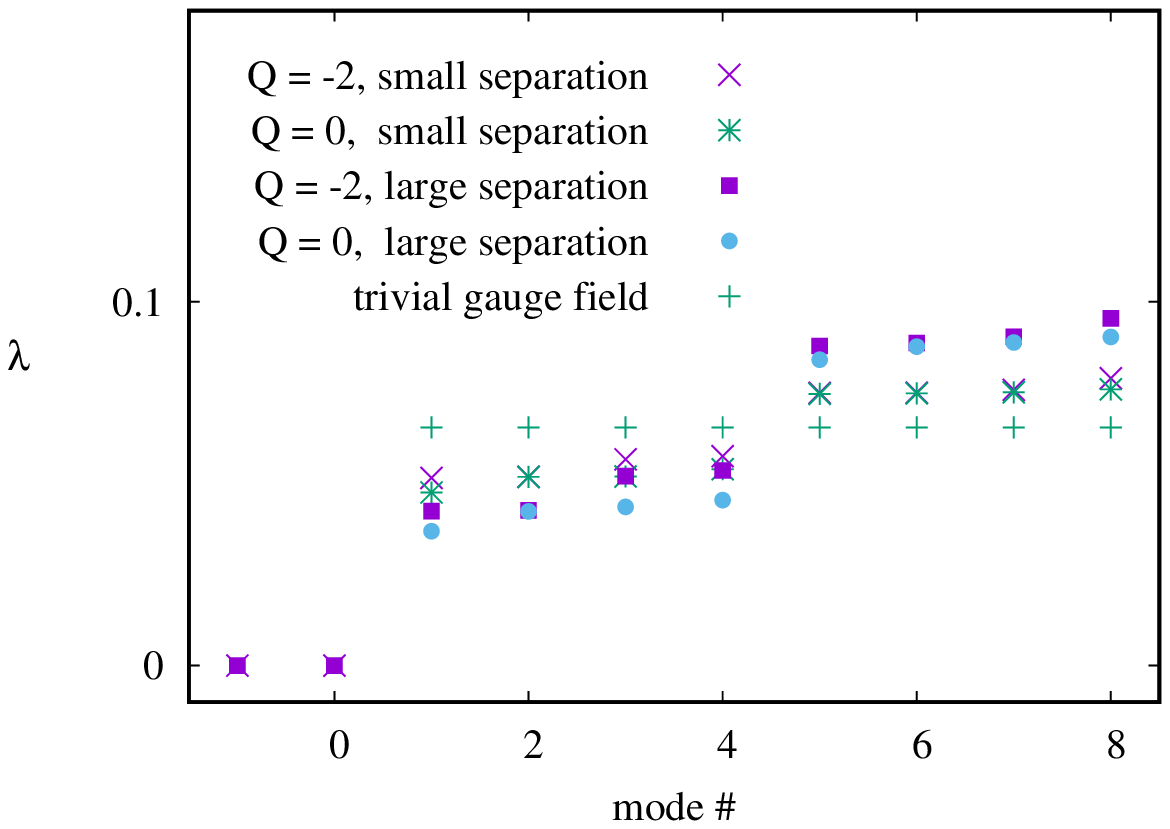}
\caption{a) The lowest overlap eigenvalues for the vortex configurations schematically displayed in Fig.~\ref{fig:schematic} with $Q=-2$ (Fig.~\ref{fig:schematic}a) and $Q=0$ (Fig.~\ref{fig:schematic}b). These values are compared with those of the free Dirac operator on a $16^4$ lattice. b) The influence of the distance between the vortex sheets is investigated for the $Q=-2$ configuration. Increasing this distance the interaction decreases and the lowest lying non-zero modes move from the lowest Matsubara frequency towards zero.}
\label{fig:overlap}
\end{figure}

As mentioned above, the topological charge of two intersecting anti-parallel vortex pairs where one of the vortices is colorful, negatively charged, leads to a total topological charge $Q=-2$ or $Q=0$ depending on the location of the colorful region. In Fig.~\ref{fig:overlap} we study the lowest eigenvalues of the overlap Dirac operator in the background of these configurations and compare them with those of the free overlap Dirac operator. For the fermions we use anti-periodic boundary conditions in temporal direction and periodic boundary conditions in spatial directions on a $16^4$-lattice. In diagram a) the parameters of the configurations are the same as those in the previous section, in diagram b) these data are compared with analogous vortex configurations where the distances between the vortices in each pair are reduced from 7 to 3 and the vortex thickness parameters d of Eq.~(\ref{eq:phi-pl0}) are correspondingly decreased from 3 to 1.5.
The topological charge of these configurations agrees with the analytical index, $\mathrm{ind}D[A]=n_--n_+=Q$~\cite{atiyah:1971rm,brown:1977bj,adams:2000rn}. For a single configuration, one never finds zero modes of both chiralities and at least one of the numbers $n_-$ or $n_+$ vanishes. As shown in Fig.~\ref{fig:overlap}a, we find two zero modes of positive chirality for the $Q=-2$ configuration and no zero-mode for $Q=0$. Note, for better comparison we indicated the two zero-modes with mode numbers $\#(-1)$ and $\#0$. By the influence of the vortices the eigenvalues occupy the space between the Matsubara frequencies, visible in the spectrum of the trivial configuration. For both non-trivial configurations four of the modes move from the first Matsubara frequency down towards zero eigenvalues. Some of them let expect to contribute in the infinite volume limit to the density of near-zero modes. In Fig.~\ref{fig:overlap}b we compare the lowest eigenvalues of configurations with decreased distances, as mentioned above, with those of the original, larger distances. With increasing distance between the lumps of topological charge we expect decreasing interaction and observe that the four lowest non-zero modes shift down from the first Matsubara frequency towards zero for both topological charges.

\begin{figure}[h!]
\centering
a)\includegraphics[scale=0.5]{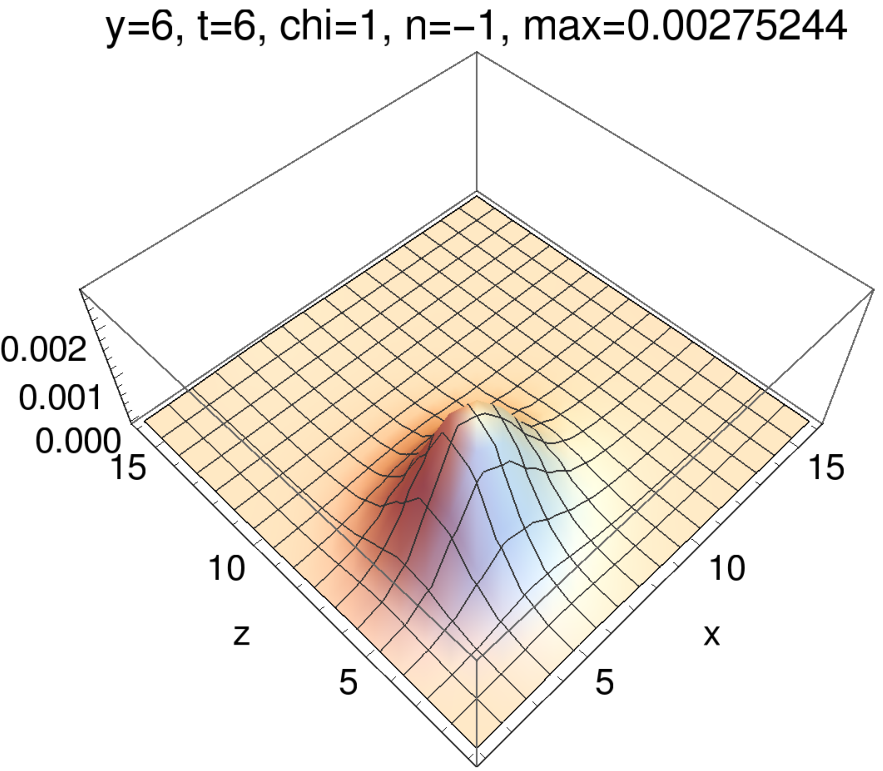}
b)\includegraphics[scale=0.5]{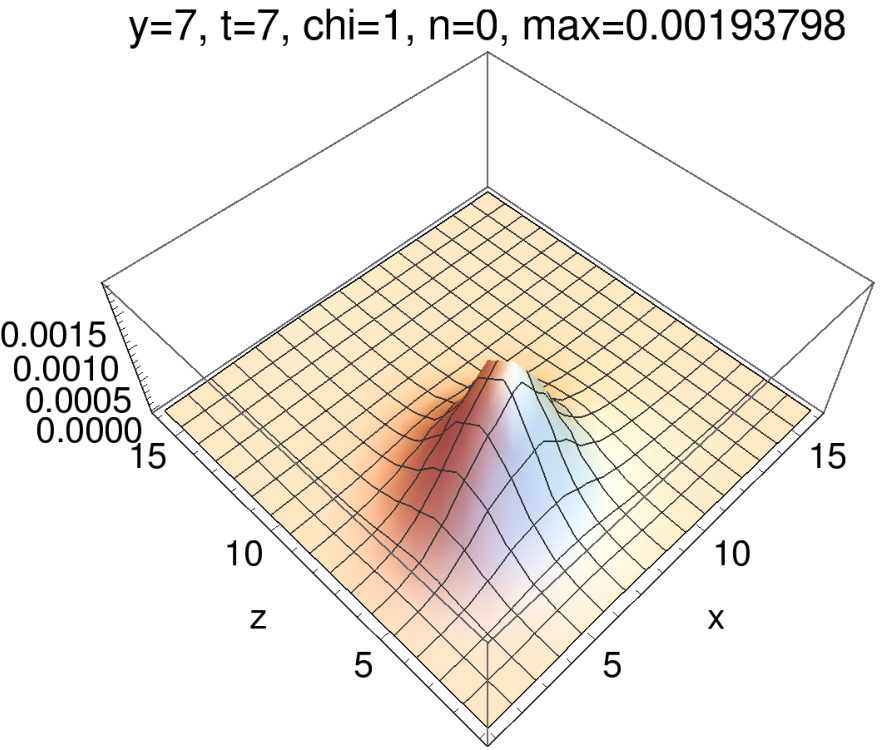}
c)\includegraphics[scale=0.5]{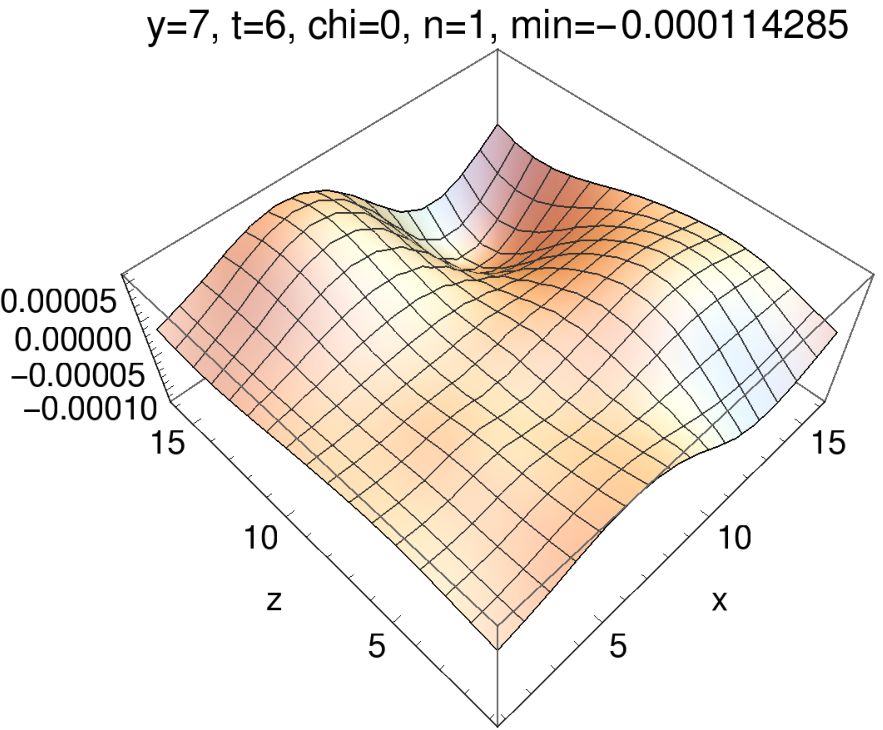}
\caption{The chiral densities $\chi_n(x)$ of the first three modes of the $Q=-2$ configuration in the x-z-planes through the points with the maximal (absolute) values of the chiral density. The plot titles indicate the plane positions, the chirality (chi=$0,\pm1$), the numbers $n=\#$ of plotted modes and the maximal (minimal) density in the plotted area, "max=..." ("min=..."). "$n=-1$" means we plot the density $\chi_{-1}(x)$ for the mode with number $\#(-1)$. Note, to ease the comparison of the various modes in Fig.~\ref{fig:overlap} we use the mode numbers $\#(-1)$ and $\#0$ for the two zero-modes and start counting positive mode numbers for non-zero modes.}
\label{fig:chirdensQm2}
\end{figure}
The two gauge field configurations with $Q=-2$ and $Q=0$ give the nice opportunity to study the properties of zero modes and near-zero modes and to compare them with excited modes. For the first one, $Q=-2$, we compare in Fig.~\ref{fig:chirdensQm2} the chiral densities of the first three modes in the x-z-planes through the points with the maximal values of the chiral density. The first two modes, $\#(-1)$ and $\#0$, are zero-modes with positive chirality. The corresponding diagrams, Figs.~\ref{fig:chirdensQm2}a and b show clear maxima, located in the intersection plane close to the center of the colorful region with topological charge $Q=-1$, slightly shifted against each other by $\Delta x_\mu=(2,1,0,1)$ lattice units. The maximum of the first mode is more pronounced than that of the second mode. The same clear peak structure we can find also in all other cross-section through the maxima. This means the peaks are nearly spherical in four dimensions. Due to the missing space we do not show the corresponding cross-sections. The next modes are non-zero modes, see Fig.~\ref{fig:overlap}a. The third mode, depicted in Fig.~\ref{fig:chirdensQm2}c looks completely different. Oscillations  extend through the whole intersection plane, but they have large amplitudes only in a region of a few lattice constant in the perpendicular y- and t-directions. The corresponding densities are again not shown. Comparing Fig.~\ref{fig:chirdensQm2}c with Fig.~\ref{fig:topden}a we see that the maxima and minima of $\chi_1(x)$ reflect the position of the topological charge density. \begin{figure}[h!]
\centering
a)\includegraphics[scale=0.56]{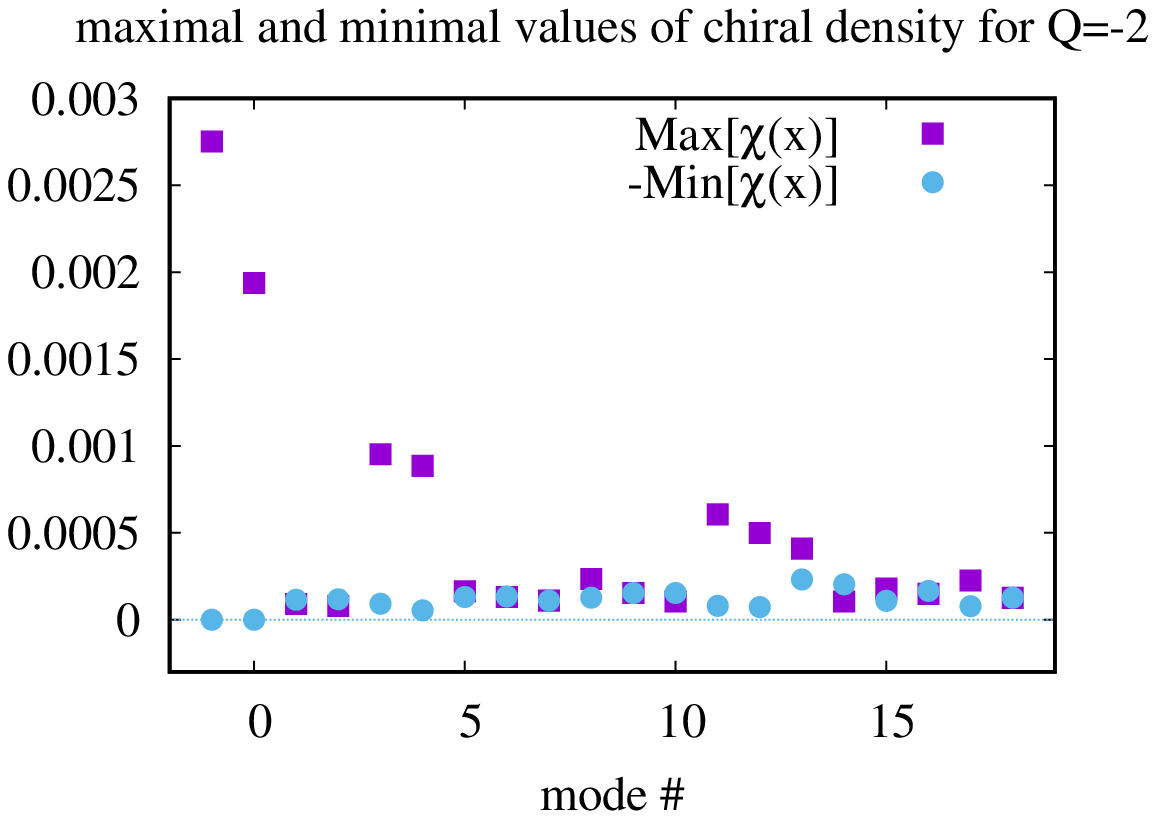}
b)\includegraphics[scale=0.56]{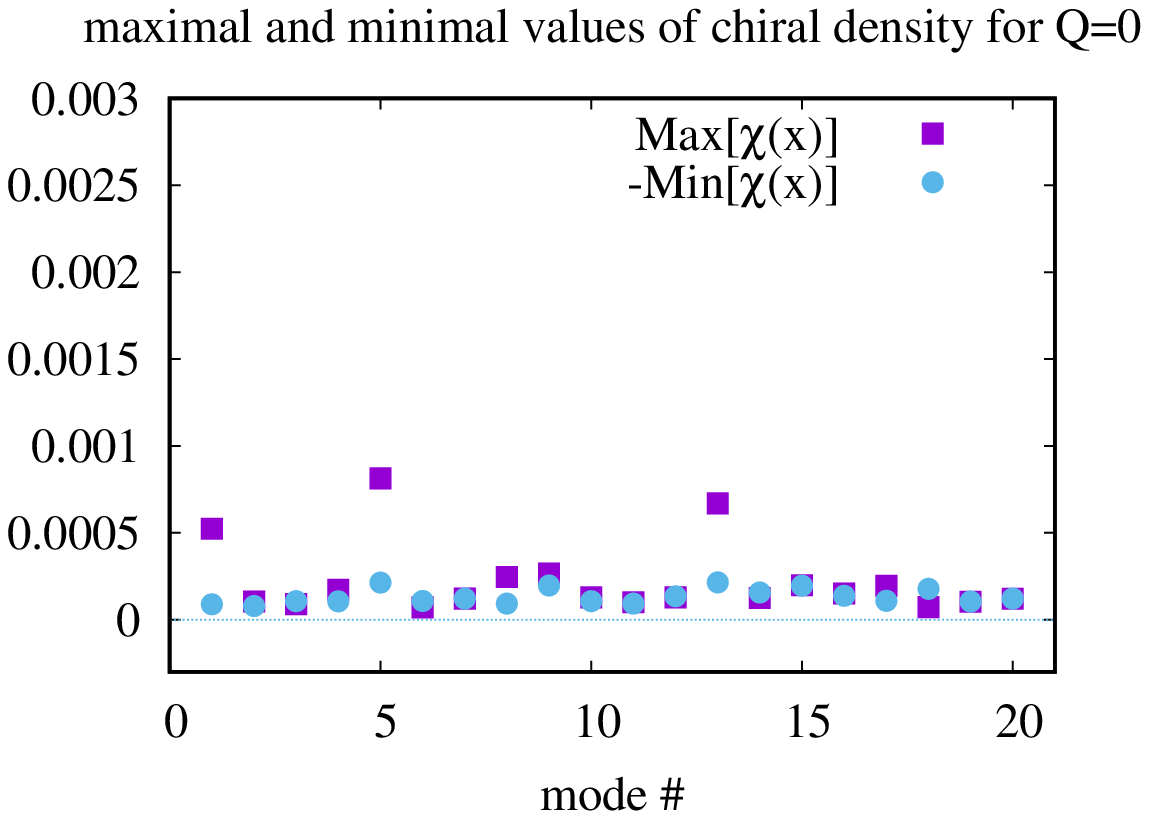}
\caption{Maximal and minimal values of the chiral densities of the 20 lowest modes for the above described configuration with $Q=-2$ (left) and $Q=0$ (right).}
\label{fig:peaks}
\end{figure}
The lumps of topological charge at the intersection points of the vortex pairs seem to contribute jointly to the shape of this mode. Lumps with positive topological charge density are correlated with the components with negative chiral density and vice versa. The chiral densities of the next modes we can only discuss in words. $\chi_2(x)$ behaves similar to $\chi_1(x)$, $\chi_3(x)$ and $\chi_4(x)$ have again positive peaks similar to $\chi_{-1}(x)$ and $\chi_0(x)$ but only around 1/3 of their height. This indicates that the maxima and minima of the topological density keep their influence on some of the next modes. In this respect it may be interesting to compare the values of maxima and minima of the chiral densities of the lowest modes.

Fig.~\ref{fig:peaks}a depicts the values of maxima and minima of the chiral densities for the 20 lowest modes for $Q=-2$ configuration. Large maxima of the density indicate localized peaks influenced by the colorful vortex region centered at $x_\mu=(6,6,6,6)$. Besides the two zero-modes such maxima can be found for the mode numbers $\#3,\#4,\#11,\#12$ and $\#13$. Inspections of these modes shows nice peaks with a height given in Fig.~\ref{fig:chirdensQm2} and a shallow negative see compensating the peak in order to reach the integrated chiral density $\chi=0$. In regions far away from the peaks there appears a wavy character due to the neighborhood of these modes to the corresponding Matsubara frequency. For points in Fig.~\ref{fig:peaks}a with nearly equal sizes of maxima and minima we can imply a wavy character due to the four vortex intersections and the increasing momenta of higher Matsubara frequencies.

Fig.~\ref{fig:peaks}b depicts maximal and minimal values of the chiral densities of the 20 lowest modes for the $Q=0$ configuration. The modes $\#1,\#5,\#13$ are striking for their large maxima and point to their peaky structure at the position of the colorful vortex around $x_\mu=(13,6,6,6)$ similar to Figs.~\ref{fig:chirdensQm2}a and b, but with lower height. These three modes are interestingly some type of band-heads in the eigenvalue spectrum of Fig.~\ref{fig:overlap}a. The other modes have wavy character, e.g. mode $\#2$ behaves similar to mode $\#1$ of the $Q=-2$ configuration displayed in Fig.~\ref{fig:chirdensQm2}c.

%------------------------------------------------------------------------------
\section{Conclusion}\label{Sect4}
In four dimensions vortices are two-dimensional surfaces with some thickness. On a periodic lattice plain vortices can only be defined in parallel or anti-parallel pairs. In the past mainly uni-color vortices were investigated. Intersections of plain uni-color vortices contribute to the topological charge with $\pm1/2$. The four intersections of two vortex pairs result therefore in topological charges of $Q=0,\,\pm1$ or $\pm2$. In this article we considered pairs of anti-parallel vortices yielding $Q=0$. It was shown recently that vortices can also have a color structure with non-vanishing topological charge. To get more insight into the effect of such color structures on vortices we investigated the influence of a circular colorful region with topological charge $Q=-1$ around one of the four intersection points of two intersecting anti-parallel vortex pairs. We studied the consequences of this insertion on topological charge density, zero-modes and near-zero modes. To uni-color vortices one can attribute an orientation. In this picture of unicolor vortices the above colorful region introduces a monopole line on its vortex surface, a line surrounding the intersection point and changes the surface orientation inside this circular region. This leads to a sign change of the topological charge contribution at the intersection. The contributions of the intersections aggregate thus to $Q=\pm1$. The monopole line itself has a non-trivial topology and leads to further contribution to the topological charge, in the considered cases with a contribution of $-1$. For our configurations we can therefore distinguish the two cases $Q=0$ and $Q=-2$. In both configurations we have identified the regions with non-vanishing contributions to the topological charge density.

Further, we have analyzed the low-lying modes of the Dirac operator. The number of zero-modes agrees with the expectations from the Atiyah-Singer index theorem. In both configurations we found four low lying modes which are shifted from the first Matsubara frequency down towards zero eigenvalues. We increased the distance between the lumps of topological charge in the expectation of decreasing interaction and observe a decrease of the eigenvalues of these four modes.

We found that the lumps of topological charge influence strongly the spatial distribution of the low-lying modes of the Dirac operator. The colorful region with topological charge $Q=-1$ leads in some of the lowest modes to distinct positive peaks of the chiral density. The other modes reflect the positions of the intersections and their contributions to the topological charge density. It turned out that a good indication for the chiral properties of the eigenmodes is the relation between maxima and minima of the chiral densities.

\acknowledgments{We would like to thank Roman H\"ollwieser and Urs M. Heller for their support in the preparation of the programs. SMHN thanks Sedigheh Deldar for her support and is grateful to the Iran National Science Foundation (INSF) for supporting this study.}

\bibliographystyle{utphys}
\bibliography{chiral}

\end{document}